\documentclass[preprint3]{aastex}                                   
\usepackage{emulateapj5}
\usepackage{fixltx2e}
\usepackage{ amsmath }
\usepackage{ amssymb }
\shorttitle{Aspect Ratio Dependence of Collapse Times}

\begin{document}
\title{ASPECT RATIO DEPENDENCE OF THE FREE-FALL TIME FOR NON-SPHERICAL SYMMETRIES}
\author{Andy Pon\altaffilmark{1,2},  Jes\'{u}s A. Toal\'{a}\altaffilmark{3,4}, Doug Johnstone\altaffilmark{2,1},  Enrique V\'{a}zquez-Semadeni\altaffilmark{4}, \\
  Fabian Heitsch\altaffilmark{5} \& Gilberto C. G\'{o}mez\altaffilmark{4}}
\altaffiltext{1}{Department of Physics and Astronomy, University of
  Victoria, P.O. Box 3055, STN CSC, Victoria, BC V8W 3P6, Canada;
  arpon@uvic.ca} 
\altaffiltext{2}{NRC-Herzberg Institute of Astrophysics, 5071
  West Saanich Road, Victoria, BC V9E 2E7, Canada;
  Douglas.Johnstone@nrc-cnrc.gc.ca}
\altaffiltext{3}{Instituto de Astrof\'{i}sica de Andaluc\'{i}a, CSIC,
  Glorieta de la Astronom\'{i}a s/n, E-18008, Granada, Spain;
  toala@iaa.es} 
\altaffiltext{4}{Centro de Radioastronom\'{i}a y Astrof\'{i}sica,
  Universidad Nacional Aut\'{o}noma de M\'{e}xico, Campus Morelia
  Apartado Postal 3-72, 58090 Morelia, Michoac\'{a}n, Mexico;
  e.vazquez@crya.unam.mx, g.gomez@crya.unam.mx} 
\altaffiltext{5}{Department of Physics and Astronomy, University of
  North Carolina Chapel Hill, CB 3255, Phillips Hall, Chapel Hill, NC
  27599, USA; fheitsch@unc.edu}

\begin{abstract}
We investigate the collapse of non-spherical substructures, such as sheets and filaments, which are ubiquitous in molecular clouds. Such non-spherical substructures collapse homologously in their interiors but are influenced by an edge effect that causes their edges to be preferentially accelerated. We analytically compute the homologous collapse timescales of the interiors of uniform-density, self-gravitating filaments and find that the homologous collapse timescale scales linearly with the aspect ratio. The characteristic timescale for an edge-driven collapse mode in a filament, however, is shown to have a square-root dependence on the aspect ratio. For both filaments and circular sheets, we find that selective edge acceleration becomes more important with increasing aspect ratio. In general, we find that lower dimensional objects and objects with larger aspect ratios have longer collapse timescales. We show that estimates for star formation rates, based upon gas densities, can be overestimated by an order of magnitude if the geometry of a cloud is not taken into account.
\end{abstract}

\keywords{ISM: clouds - ISM: structure - stars: formation}

\section{INTRODUCTION}
\label{introduction}

Molecular clouds are observed to have complex geometries and contain non-spherical substructures, including sheets and filaments (e.g., \citealt{Schneider79, Bally87, Lada99, Hartmann02, Johnstone03, Lada07, Myers09, Molinari10, Andre10}). Recent {\it Herschel} observations, in particular, have revealed a plethora of filamentary structures within star-forming regions (e.g., \citealt{Andre10}). Filamentary structures are seen on both molecular cloud scales as well as on the small scales of individual protostellar envelopes (e.g., \citealt{Andre10, Tobin10, Hacar11}). Such filamentary structures are also commonly predicted by star formation models and formed in molecular cloud simulations, although their formation mechanisms vary depending upon which model is used. Supersonic hydrodynamic and magnetohydrodynamic turbulence, gravo-turbulent models, gravitational amplification of anisotropies, flow collisions, and global gravitational accelerations have all been shown to be capable of forming filamentary structures (e.g., \citealt{Lin65, Klessen01, Padoan01, Burkert04, Hartmann07, Padoan07, Bate09, VazquezSemadeni10, VazquezSemadeni11}).

The collapse modes of spherical and infinite, non-spherical structures have been thoroughly investigated in earlier works (e.g., \citealt{Ledoux51, Stodolkiewicz63, Ostriker64, Larson69, Penston69II, Shu77,Larson85, Inutsuka92, Nakamura93, Inutsuka97, Fiege00, Curry00, Myers09}), but the collapse properties of finite, non-spherical structures have only been considered recently and are much less understood, owing to their inherent global and local instabilities, as well as the critical importance of initial conditions (e.g., \citealt{Bastien83, Bastien91, Burkert04, Hsu10, Pon11, Toala12}).

Simulations show that non-spherical structures collapse on longer timescales than equal-density spherical objects (e.g., \citealt{Burkert04, VazquezSemadeni07}) and that non-spherical structures are prone to gravitational focusing, whereby strong density enhancements form at the edges of these objects (e.g., \citealt{Bastien83, Burkert04, Hartmann07, Heitsch08Slyz, Hsu10}). The longer collapse timescales of non-spherical objects, as well as the presence of selective edge acceleration, have also been analytically demonstrated \citep{Burkert04, Pon11, Toala12}.

While uniform-density spheres collapse homologously (e.g., \citealt{Binney87}), an edge effect in circular sheets and filaments produces a second mode through which these lower dimensional clouds can collapse, wherein the collapse is controlled by an infalling edge sweeping up material. It has not been clear, however, whether the edge of a lower dimensional structure will obtain sufficient momentum to be able to sweep up the interior, so that the collapse timescale is controlled by the selective edge acceleration, or whether the roughly homologously collapsing interior will slow down the collapsing edge enough such that the collapse timescale will approach the homologous collapse timescale of the interior.

Recently, \citet{Toala12} showed that the free-fall times of circular sheet-like (``2D'' collapse) and filamentary clouds (``1D'' collapse) depend strongly on the geometry of the cloud and, for both cases, are larger than that of a uniform sphere with the same volume density by a factor proportional to the square root of the initial aspect ratio, $A$. For circular sheets, the aspect ratio is given by $A=R(0)/H$, $R(0)$ being a sheet's initial radius and $H$ its thickness, and for filaments, the aspect ratio is defined as $A=Z(0)/{\cal R}$, where $Z(0)$ is a filament's initial half-length and ${\cal R}$ is the radius. On the other hand, \citet{Pon11} find that the collapse timescale of an infinitely thin filament varies linearly with the aspect ratio, rather than with the square root of the aspect ratio.

\citet{Toala12} obtain their analytic expressions for the free-fall times by integrating, over time, the accelerations at the edges of a circular sheet and cylinder under two different assumptions. In the first case, \citet{Toala12} assume that the mass of the cloud remains constant and that the density of the cloud remains spatially uniform over the entire collapse. This is equivalent to assuming that the collapse is homologous. In their second case, \citet{Toala12} assume that the density of the cloud remains spatially and temporally constant. For this second case, it is assumed that the collapse is dominated by the insweeping edge. These approximations are used to make the problem analytically tractable. \citet{Toala12} argue that these two approximations represent the extreme cases, where the collapse timescale is determined solely from either the homologously collapsing interior or from the selective edge acceleration, such that the actual collapse timescale should lie somewhere between the two derived timescales.

In Section \ref{homologous}, we calculate the collapse timescales of a uniform-density cylinder, under the assumption that the cylinder collapses homologously along its major axis. We use the first-order approximation to the acceleration that is valid in the interior of the cylinder, rather than using the acceleration at the edge, as done by \citet{Toala12}. In Section \ref{uniform density}, we calculate the collapse timescale for a filament, under the approximation that the interior density remains spatially and temporally constant, by treating the gravitational force per unit mass on the edge as a rate of momentum transfer per unit mass, rather than as an acceleration, as done by \citet{Toala12}. We thus take into account the effect of low velocity mass being accumulated by the edge. We compare our collapse timescales to previously obtained results in Section \ref{comparison}. In Section \ref{discussion}, we compare the relative importance of the homologous collapse mode and edge-driven collapse mode in filaments and circular sheets. We also discuss in Section \ref{discussion} whether the collapse timescales of filaments and circular sheets have the same dependence upon the aspect ratio, as found by \citet{Toala12}, as well as discussing the implications of our collapse timescales. Finally, we summarize our findings in Section \ref{conclusion}.

\section{COLLAPSE TIMESCALES OF UNIFORM DENSITY CYLINDERS}
\label{1D}

While circular sheets and cylinders are formally the same type of object, we differentiate between the two based upon which axis is longer and along which axis collapse is occurring. We refer to objects with radii larger than their heights, and collapsing radially, as finite, circular sheets and refer to objects with heights larger than their radii, and collapsing along their long, vertical axis, as cylinders. Thus, finite circular sheets represent ``2D'' sheets while cylinders represent ``1D'' filaments. 

\subsection{Homologous Collapse}
\label{homologous}

The free-fall collapse timescale of a sphere is a well-studied problem (e.g., \citealt{Binney87}) and the homologous collapse timescale of the interior of a finite circular sheet is calculated by J. A. Toal\'{a} et al.\ (2012, in preparation [erratum]). For reference, we present derivations of the collapse timescales of these two objects in Appendices \ref{3d} and \ref{2d}.

We examine a cylinder with a total mass $M$, a total length along the major axis of $2Z(t)$, and a volume density $\rho(t) = M/ [2 \pi {\cal R}^{2} Z(t)]$, where ${\cal R}$ is the time-independent radius. We denote the distance of a mass element along the major axis from the center of the cylinder as $z$, the initial length of the cylinder as $2Z(0)$, and the initial volume density as $\rho(0)$. The aspect ratio of the cylinder is defined as $A = Z(0) / {\cal R}$. The magnitude of the acceleration along the major axis of the cylinder, for points within the cylinder, is given by \citet{Burkert04} as
\begin{eqnarray}
  a(z, t) &=& 2\pi G \rho(t) \big\{2z +\sqrt{{\cal R}^2+(Z(t) - z)^2}\nonumber \\
  &&-\sqrt{{\cal R}^2+(Z(t)+z)^2}\big\}.
\label{eqn:a1dexact}
\end{eqnarray}

In a homologous collapse, the density remains spatially uniform, although not temporally constant, and all regions have the same collapse timescale. For a uniform-density object to collapse homologously, the acceleration across the object at any given time must be a linear function of the radial distance to the collapse center. Equation (\ref{eqn:a1dexact}) is not a linear function of $z$ and thus, the collapse of a filament is not homologous. \citet{Burkert04} find that the first-order approximation to the acceleration\footnotemark, under the condition that $|Z(t) - z| \gg {\cal R}$, is
\begin{equation}
  a(z,t) \approx \pi G {\cal R}^{2} \rho(t) \left[\frac{2z}{Z(t)^2 - z^2}\right].
\label{eqn:burkertapprox}
\end{equation}
While Equation (\ref{eqn:burkertapprox}) is also not a linear function of $z$, for the interior of a filament, where $z^2 \ll Z(t)^2$, the acceleration can be approximated by
\begin{equation}
  a(z,t) \approx \pi G {\cal R}^{2} \rho(t) \left[\frac{2z}{Z(t)^2}\right].
\label{eqn:a1dhomo}
\end{equation}

\footnotetext{Equation (11) of \citet{Burkert04} gives another form of this first-order approximation, but is missing a negative sign from the first term in the brackets. In our notation, \citeauthor{Burkert04}'s \citeyearpar{Burkert04} Equation (11) should read as $|a(z,t)| \approx \pi G \, \rho(t) R^2 \left[-\left(Z(t)+z\right)^{-1} + \left(Z(t) -  z\right)^{-1}\right]$.}

Since Equation (\ref{eqn:a1dhomo}) is a linear function of $z$, the assumption of homologous collapse is reasonable for the interior portion of a cylinder. As shown in Appendix \ref{2d}, this is also the case for the interior of a circular sheet. While strictly Equation (\ref{eqn:a1dhomo}) should only be used to find accelerations of the interior of a cylinder, the collapse timescales in a homologous collapse are constant across the entire object and thus, evaluating this acceleration at the edge will yield the collapse timescale of the interior. In this paper, the ends of the major axis of a cylinder are referred to as the edges of the cylinder.

Equation (\ref{eqn:a1dhomo}) can be re-written in terms of the total mass of a cylinder. Evaluating this new expression at the edge, where $z = Z(t)$, yields
\begin{equation}
\frac{dv_0(t)}{dt} = \frac{G M}{Z(t)^2},
\label{eqn:1dode}
\end{equation}
where $v_0(t)$ is the velocity at the edge at time $t$. This differential equation has the same dependences on mass and length as in the spherical and circular sheet cases presented in Appendices \ref{3d} and \ref{2d}. Equation (\ref{eqn:1dode}) can be solved for the cylinder collapse timescale, $\tau_\mathrm{1D}$,
\begin{equation}
  \tau_\mathrm{1D} = \sqrt{\frac{2}{3}} \, A \, \tau_\mathrm{3D},
\end{equation}
where $\tau_\mathrm{3D}$ is the classical free-fall timescale of a uniform-density sphere with the same volume density as the cylinder, as derived in Appendix \ref{3d}. Thus, the homologous collapse timescale for a cylinder is linearly proportional to the aspect ratio.

\subsection{Cylindrical Edge Collapse: A Constant Density Approximation}
\label{uniform density}

The exact solution for the accelerations of a cylinder, Equation (\ref{eqn:a1dexact}), shows that nonlinear terms become significant toward edges, such that the edges are preferentially given more momentum than would be expected for homologous collapse. If edges are given sufficient momentum, the collapse of a cylinder may be dominated by the edges sweeping up interior material, such that the collapse occurs on a timescale faster than the homologous collapse timescale calculated in Section \ref{homologous}.

We now calculate the collapse timescale of a cylinder under the approximation that the material inside of the edge stays at a constant density and does not move inward until contacted by the insweeping edge. We assume that the only acceleration in the system is due to the gravitational force on the edge caused by the interior material and we assume that the edge sweeps up all material it contacts, such that the mass of the edge grows as it falls inward.  We do not consider the gravitational force on one edge of the filament due to the other edge. For a cylinder with an aspect ratio of 10, the gravitational force due to the second edge only becomes equal to the force due to the uniform-density, interior material once the cylinder has shrunk to roughly one tenth of its original size. Thus, the collapse timescale is relatively unaffected by the presence of a mass concentration at the other edge. As argued by \citet{Toala12}, the true collapse timescale should lie between the homologous collapse approximation and this constant density approximation.

As given by \citet{Pon11} and \citet{Toala12}, the gravitational force per unit mass on the major axis and at the edge of a cylinder of length $2Z$, radius ${\cal R}$, and uniform-density $\rho$ is
\begin{equation}
g = 2\pi \, G \, \rho \left[2Z(t) + {\cal R} - \sqrt{{\cal R}^2 + 4Z(t)^2}\right].
\label{eqn:a1dendexact}
\end{equation}

Under the assumption that $Z(t) \gg {\cal R}$, the square root can be expanded and the force per unit mass, to lowest order, becomes
\begin{equation}
g \approx 2\pi \, G \, \rho \, {\cal R}.
\end{equation}
Note that the force per unit mass is independent of the length of the cylinder. This approximation to the acceleration at the end of a cylinder is within 5\% of the exact acceleration given by Equation (\ref{eqn:a1dendexact}) for aspect ratios above 5 and within 10\% for an aspect ratio of 3.

By equating the gravitational force acting on the edge with the total rate of change of momentum of the edge, we find, in Appendix \ref{app:filament dens}, that the length of the cylinder at a time $t$ is given by
\begin{equation}
Z(t)  \approx Z(0) - \frac{g t^2}{6},
\label{eqn:zoft1d}
\end{equation}
where $Z(0)$ is the initial length of the cylinder. 

Equation (\ref{eqn:zoft1d}) shows that the effect of the edge of a cylinder accreting low velocity mass is only to lower the effective acceleration by a factor of three below the acceleration that would be obtained by directly equating the gravitational force per unit mass to the acceleration of the edge. Substituting in the lowest order approximation to the force per unit mass at the edge yields a collapse timescale of
\begin{eqnarray}
\tau_{\mathrm{1D}} &=& \sqrt{\frac{6 Z(0)}{2\pi \, G \, \rho \, {\cal R}}},\\
\tau_{\mathrm{1D}} &=& \sqrt{\frac{32 A}{\pi^{2}}} \, \tau_\mathrm{3D},
\end{eqnarray}
where, as before, $A$ is the original aspect ratio and $\tau_\mathrm{3D}$ is the classical free-fall timescale of a uniform-density sphere with the same volume density as the cylinder.
  
\subsection{Comparison to Previous Works}
\label{comparison}

\citet{Pon11} examine the accelerations of an infinitely thin, but finitely long, filament and find that the acceleration becomes infinite at the edge of such a filament. Because of this infinite acceleration, \citet{Pon11} derive a first-order approximation to the accelerations of their infinitely thin filament that is identical to the first-order approximation to the accelerations of a finite radius cylinder derived in Section \ref{1D} as Equation (\ref{eqn:a1dhomo}). 
 
We find that the homologous collapse timescale for a cylinder is $\sim0.82 A \, \tau_\mathrm{3D}$. This differs significantly from the timescale found by \citet{Toala12} for the homologous collapse of a cylinder, $\sim0.92\sqrt{A} \, \tau_\mathrm{3D}$. We find that the collapse timescale for a homologously collapsing cylinder scales linearly with the aspect ratio, while \citet{Toala12} find a $\sqrt{A}$ relationship. The cause of the difference between these two results is that \citet{Toala12} use the acceleration at the edge, while we use a first-order approximation to the acceleration in the interior of a cylinder. Thus, we probe the collapse timescales of the interior of a cylinder while \citet{Toala12} are sensitive to the collapse dynamics at the edge of a cylinder.

We find that if the collapse of a cylinder is dominated by the preferential edge acceleration, such that the interior remains static and at a constant density until swept up by the edge, the collapse timescale is $\sqrt{32 A/\pi^{2}} \, \tau_\mathrm{3D}$. \citet{Toala12} find that the collapse timescale of a cylinder, under this constant density approximation, is $\sqrt{32 A / (3\pi^{2})} \, \tau_\mathrm{3D}$. Since we show in Section \ref{uniform density} that the effect of accounting for the additional mass being accreted by the edge of the cylinder is to decrease the effective acceleration by a factor of three, it is unsurprising that the \citet{Toala12} collapse timescale is exactly a factor of $\sqrt{3}$ smaller than what we derive.

\section{DISCUSSION} 
\label{discussion}

\begin{figure*}[htbp] 
   \centering
   \includegraphics[width=5.5in]{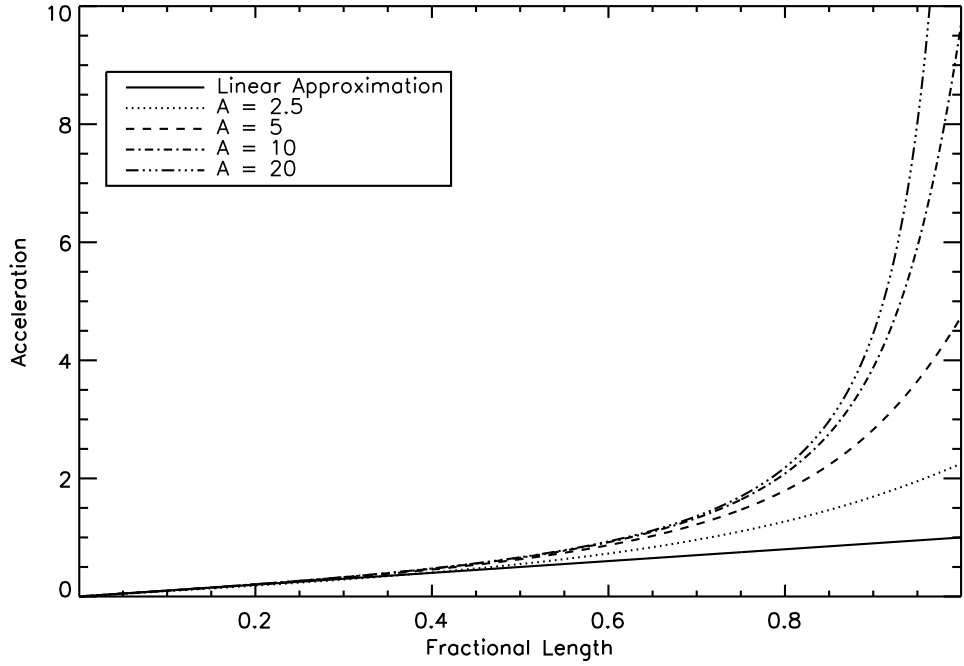}
   \caption{Radial accelerations, in units of $2 \lambda G / Z$, for uniform-density cylinders with different aspect ratios. The dotted, dashed, dash-dotted, and dash-triple-dotted lines show the exact radial accelerations for cylinders with aspect ratios of 2.5, 5, 10, and 20, respectively. The solid line shows the first-order approximation to the accelerations for all four cylinders. Note how the assumption of homologous collapse becomes worse as the aspect ratio increases.}
      \label{fig:aprofile1d}
\end{figure*}

Figure \ref{fig:aprofile1d} shows the exact accelerations, from Equation (\ref{eqn:a1dexact}), of uniform-density cylinders with various aspect ratios. The aspect ratios shown range from 2.5 to 20 by factors of two. The units of acceleration in Figure \ref{fig:aprofile1d} are $2 \lambda G / Z$, where $\lambda = \pi {\cal R}^2 \rho$, such that the linear approximations to the accelerations of all of the cylinders are given by the same solid line. The momentum deposited per unit length, per unit time of a cylinder can be found by multiplying the acceleration by a constant factor and thus, is not shown in Figure \ref{fig:aprofile1d}.

\begin{figure*}[htbp] 
   \centering
   \includegraphics[width=5.5in]{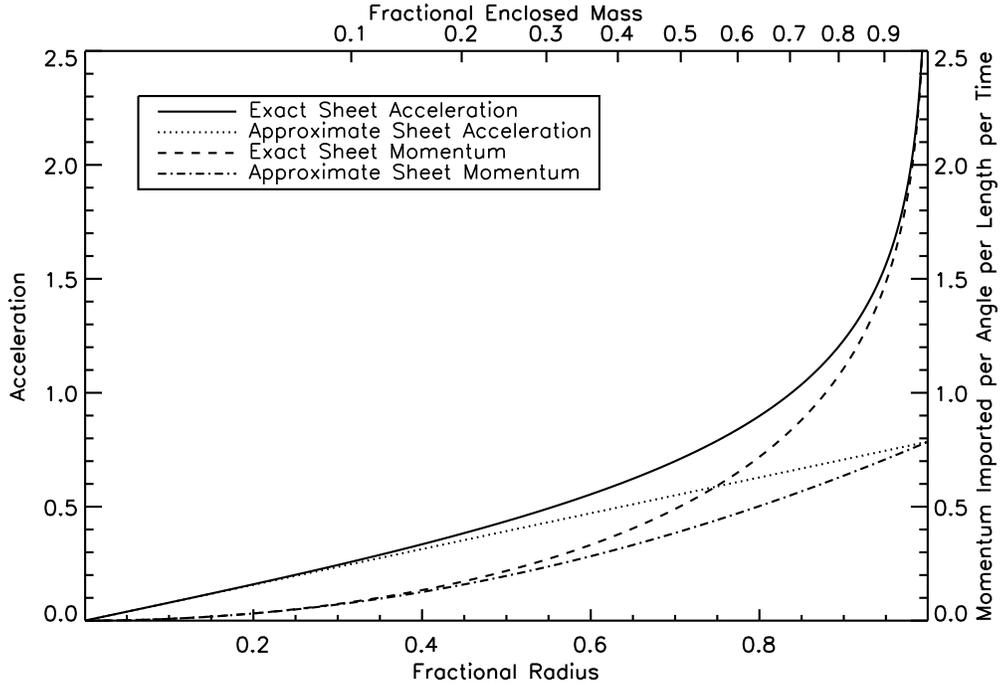}
   \caption{Exact radial accelerations, in units of $4G \Sigma$, for an infinitely thin, uniform-density, circular sheet are shown as the solid line. The first-order approximation to these accelerations is shown as the dashed line. The dotted and dash-dotted lines show the momentum imparted per unit angle, per unit radial length, per unit time, in units of $4G \, \Sigma^2 R$, based upon the exact and approximate accelerations, respectively. The lower $x$-axis shows the distance along the sheet as a fraction of the total radius and the upper $x$-axis gives the fraction of the total mass of the sheet within that radius.}
      \label{fig:aprofile2d}
\end{figure*}

The exact accelerations of a radially collapsing, infinitely thin, unifor-density, circular sheet with mass $M$, radius $R(t)$, and surface density $\Sigma(t)$, are given by Equation (\ref{eqn:a2dexact}) in Appendix \ref{2d}, while the first-order approximations to these accelerations are given by Equation (\ref{eqn:a2dapprox}). Figure \ref{fig:aprofile2d} shows these exact and first-order accelerations of a circular sheet in units of $4G\Sigma$. Figure \ref{fig:aprofile2d} also shows the total momentum deposited per unit angle, per unit radial length, per unit time of a circular sheet, as a function of enclosed mass, for these two different acceleration equations, in units of $4G\Sigma R$. The momentum deposited per unit angle, per unit radial length, per unit time, is related to the acceleration via
\begin{equation}
p(r, t) = a(r, t) r.
\end{equation}

Figures \ref{fig:aprofile1d} and \ref{fig:aprofile2d} show that, for the interiors of uniform-density cylinders and circular sheets, the accelerations can be well approximated by linear functions of distance. A comparison of Equations (\ref{eqn:a1dexact}) and (\ref{eqn:a2dexact}) to Equations (\ref{eqn:a1dhomo}) and (\ref{eqn:a2dapprox}) shows that this linear approximation is valid to within a factor of two for $z < 0.7 Z$ in a uniform-density cylinder with an aspect ratio of 10 and for $r < 0.94 R$ in a uniform-density circular sheet. Near the edges of cylinders and circular sheets, nonlinear terms become dominant and the accelerations become significantly larger than the first-order approximations. This suggests that the collapse of cylinders and circular sheets can be described as the combination of two separate collapse modes. The interiors of these objects collapse roughly homologously while the outer regions are dominated by an edge-driven collapse mode, wherein momentum is preferentially deposited toward the edges of the objects and collapse proceeds as the edges sweep up slower moving interior gas. This preferential edge acceleration has been previously identified in simulations and analytic work (e.g., \citealt{Burkert04, Hartmann07, Pon11}).

We associate the collapse timescales of Sections \ref{homologous} and \ref{uniform density} with the two different collapse modes. We consider the homologous collapse timescale derived in Section \ref{homologous} to be the characteristic collapse timescale of the interior of a cylinder and the constant density collapse timescale derived in Section \ref{uniform density} to be the characteristic collapse timescale for the edge-driven collapse mode in a cylinder.

The different dependences on aspect ratio, for a cylinder's collapse timescale, as derived in Sections \ref{homologous} and \ref{uniform density}, is thus explainable as being due to the different calculations probing different collapse modes. Our results suggest that the interior of a cylinder will collapse on a timescale proportional to $A$ while the edge-driven collapse timescale will depend on $\sqrt{A}$. Furthermore, this suggests that the relative importance of the preferential edge acceleration will also depend upon the aspect ratio, with the edge collapse mode being more important in cylinders with larger aspect ratios. Such a trend is clearly seen in Figure \ref{fig:aprofile1d}.

\subsection{Relative Importance of Preferential Edge Acceleration versus Homologous Collapse}
\label{relative}

While the actual collapse of a cylinder or a circular sheet will be a combination of the homologous collapse mode and the edge-driven collapse mode, it is possible that one of these modes will be dominant. If the edge-driven collapse mode is dominant, such that the majority of the total momentum is injected at the edge, the collapse will proceed primarily by the edge falling in and sweeping up material. Alternatively, if the majority of the momentum imparted to an object is due to the linear term of the acceleration, the infalling edge will have insufficient momentum to significantly accelerate the material it sweeps up and the collapse will proceed roughly homologously, albeit with a slight density enhancement at the edge.

The total rate of momentum imparted to a circular sheet can be calculated by integrating the product of Equation (\ref{eqn:a2dexact}) and the mass element over the entire sheet or by multiplying the total area under the dashed curve in Figure \ref{fig:aprofile2d} by 2$\pi$. The total momentum imparted to a cylinder can be found by integrating Equation (\ref{eqn:a1dexact}) over the entire cylinder and then multiplying by the total mass of the cylinder, or by multiplying the area under the appropriate curve in Figure \ref{fig:aprofile1d} by the mass of the cylinder. 

For an infinitely thin circular sheet, the total rate of momentum imparted to the sheet is finite, whereas for an infinitely thin filament, the total rate of momentum imparted is infinite. Thus, there is a maximum limit to the fractional contribution of the nonlinear components of the acceleration to the total momentum imparted to a circular sheet. For an infinitely thin circular sheet, almost twice as much momentum is imparted to the sheet due to the linear term in the acceleration than due to the nonlinear terms. Since the importance of preferential edge acceleration shrinks with decreasing aspect ratio, the linear acceleration term will dominate the momentum imparted for all circular sheets with finite height. Thus, it is expected that circular sheets should collapse roughly homologously.

\citet{Burkert04} simulate the collapse of uniform-density circular sheets and find that the collapse timescales of the sheets are only 20\% shorter than the homologous collapse timescale that we calculate. This shortening of the timescale by 20\% is almost exactly what would be expected given that nonlinear terms contribute an additional 50\% to the momentum of infinitely thin circular sheets. \citet{Burkert04} also note that the interiors of their simulated circular sheets undergo significant collapse before encountering the edge, as would be expected only if the homologous collapse mode is significant in these circular sheets.

For uniform-density cylinders with aspect ratios larger than five, the nonlinear components of the acceleration impart more momentum to the cylinders than the linear component. The nonlinear components contribute relatively more momentum as the aspect ratio increases and by an aspect ratio of 10, the nonlinear components contribute almost twice as much momentum as the linear component. Since realistic filaments have aspect ratios up to 60 \citep{Andre10}, preferential edge acceleration may control the evolution of many observed filaments.

\subsection{Interpretation}
Equations (\ref{eqn:3dode}), (\ref{eqn:2dode}), and (\ref{eqn:1dode}) are the differential equations describing the collapse of spheres, circular sheets, and cylinders, respectively, under the assumption of homologous collapse, and all three equations are of the form
\begin{equation}
 \frac{dv_{0}}{dt} \sim \frac{GM}{\chi^{2}},
\end{equation}
where $\chi$ is the collapsing dimension. Because of the similarity in these differential equations, the collapse timescales of the three different objects are all of the form
\begin{equation}
\tau \sim \sqrt{\frac{\pi^{2} \chi^{3}}{8 G M}}.
\end{equation}
When written in terms of the total mass of the cloud, the collapse timescales for all three objects are independent of the initial aspect ratio. This lack of dependence upon the aspect ratio is not trivial, as dimensional arguments place no constraints on the proportionality of the unitless aspect ratio. 

The dependence of the homologous collapse timescales, of spheres, circular sheets, and cylinders, on the aspect ratio, when written in terms of the initial density, comes solely from the conversion between total mass and initial density. Thus, the proportionality of the aspect ratio naturally changes with the changing dimensionality of the initial cloud. Each reduction of dimension produces an additional $\sqrt{A}$ dependence in the collapse timescale. Since, by definition, $A > 1$, the ``2D'' and ``1D'' collapse timescales are larger than that of a sphere with the same volume density. Similarly, objects of the same dimensionality but with larger aspect ratios also take longer to collapse. The increasing importance of an edge collapse mode with increasing aspect ratio, in cylindrical structures, will partially reduce the difference in collapse times between cylinders with different aspect ratios, as well as the difference between circular sheets and cylinders. Such an edge-driven collapse, however, will still occur on timescales longer than the corresponding spherical collapse timescale due to the $\sqrt{A}$ dependence that the edge-driven collapse timescale has.

\subsection{Implications}
\label{implications}

As discussed by \citet{Toala12}, the spherical free-fall timescale, $\tau_\mathrm{3D}$, is often used to calculate collapse timescales, and thus star formation rates, from observed gas densities, regardless of geometry. Since recent observations reveal a multitude of non-spherical substructures within molecular clouds (e.g., \citealt{Myers09, Molinari10, Andre10}), using the spherical free-fall timescale underestimates collapse timescales and overestimates star formation rates. In particular, attempts to predict the total galactic star formation rate from the observed gas properties of the Milky Way, by dividing the total molecular mass of the galactic interstellar medium by the spherical free-fall time corresponding to the mean density and temperature of the molecular gas, produce values of at least $30 \mbox{ }M_\odot \mbox{ yr}^{-1}$ (e.g., \citealt{Zuckerman74Palmer, Zuckerman74Evans}), while more direct, observational determinations of the galactic star formation rate, based upon emission from and number counts of young, massive stars, yield star formation rates closer to a few solar masses per year (e.g., \citealt{Smith78, Diehl06, Misiriotis06, Murray10, Robitaille10}).

For a filamentary structure with an aspect ratio of 60, corresponding to the upper limit of observed filamentary aspect ratios \citep{Andre10}, the homologous collapse timescale is a factor of almost 50 slower than the corresponding spherical free-fall time. The edge-driven collapse timescale, while faster than the homologous collapse timescale, is still almost 15 times slower than the spherical free-fall time. For a more typical aspect ratio of 10, both the homologous and edge-driven collapse timescales of such a filament would be roughly seven times slower than the spherical free-fall timescale. Thus, geometric considerations can account for a considerable portion of the discrepancy between observed and predicted galactic star formation rates, although they are unlikely to account for the entire discrepancy. For a further discussion on the implications of ``1D'' and ``2D'' collapsing objects having longer timescales, please see \citet{Pon11} and \citet{Toala12}.

The two collapse modes studied in this paper, the homologous and edge-driven collapse modes, are both global collapse modes. That is, these modes cause a cloud to collapse into one central object. Density perturbations within molecular clouds will introduce local collapse modes and these local collapse modes must operate on timescales less than the global collapse modes, as molecular clouds are observed to fragment into clusters of stars, rather than collapsing to form million solar mass stars.

\citet{Pon11} examine the conditions under which local collapse modes are significantly faster than the homologous global collapse mode in spheres, circular sheets, and cylinders. They find that strong perturbations are required for local collapse modes to be significantly faster in circular sheets and spheres, but small ($\sim$$10\%$) density perturbations in a cylinder can collapse significantly (three times) faster than the entire cylinder if the total length of the cylinder is greater than 10 times the length of the perturbation. Since thermal motions support perturbations smaller than the Jeans length and the radial length scale of a radially supported cylinder is also approximately the Jeans length \citep{Stodolkiewicz63, Ostriker64}, \citet{Pon11} find that local collapse modes are most effective in high aspect ratio cylinders. Unfortunately, it is in these large aspect ratio cylinders that the preferential edge acceleration produces significantly faster collapse timescales than the homologous collapse timescale. Thus, cylinders, as well as circular sheets and spheres, may require strong density perturbations or large-scale support mechanisms for local collapse modes to be significantly faster than global collapse modes. Note, however, that preferential edge acceleration naturally produces strong density perturbations along the edges of circular sheets and filaments, as seen in simulations of both circular sheets (e.g., \citealt{Burkert04}) and filaments (e.g., \citealt{Bastien83}).

\subsection{Caveats}
\label{caveats}
		
Realistic cloud structures are not perfect uniform-density spheres, circular sheets, nor filaments, and deviations from such perfect, symmetrical shapes will influence the collapse properties of the clouds. For instance, \citet{Burkert04} show, via simulations, that deviations from axisymmetry produce gravitational focusing, whereby local density enhancements are readily formed.

For the timescales calculated in this paper, it is assumed that clouds have sharp density boundaries. More realistic clouds are likely to taper off slowly at the edges. The introduction of such density tapers at the edges of clouds is known to reduce the significance of the preferential edge acceleration, although such a taper has to be quite large in comparison to the size of the constant density interior before the preferential edge acceleration is significantly weakened \citep{Nelson93, Li01, Pon11}. As such, preferential edge acceleration may be slightly weaker in realistic clouds than assumed here, but edge-driven collapse modes should still be important for reasonably elongated filaments.

Star-forming regions are observed to be turbulent and this effect can provide support against gravitational collapse (e.g., \citealt{Hennebelle11, Padoan11}). Thermal pressure, rotation, and magnetic fields can also significantly alter the collapse of a cloud. Thermal pressure, however, is only effective at supporting objects on scales smaller than the Jeans length, whereas observed filaments are often much longer than the Jeans length (e.g., \citealt{Andre10}). \citet{Burkert04} also point out that, if solid body rotation were supporting the interior of non-spherical clouds, the edge would still collapse due to the preferential edge acceleration, and if the exterior were rotationally supported, the central regions would be moving too rapidly and would expand. Finally, magnetic fields are only capable of providing support perpendicular to the field lines and thus, may not be capable of supporting filaments and circular sheets depending upon the orientation of the magnetic field. Models of filament formation predict that magnetic fields can be either parallel or perpendicular to the long axis of a filament \citep{Nagai98} and observational studies have found magnetic fields that are both perpendicular and parallel to the long axes of filamentary structures \citep{Goodman90, Houde04, Vallee06, Vallee07, Schneider10Csengeri, Chapman11, Sugitani11}.

The influence of collapse along the short axis of any cloud has not been considered because thermal support will generally be more effective at preventing collapse along shorter axes. In deriving the collapse timescales, the assumption that $A \gg 1$ has been utilized. While this assumption clearly breaks down at late times in the collapse, the collapse timescale should be primarily dependent upon the early stages of the collapse when infall velocities are still relatively small. Local collapse modes are also expected at later times in the collapse when, as a consequence of the density increase caused by the global collapse, the short dimension becomes larger than the Jeans length. 

The cases considered in this paper, pure homologous collapse and pure edge-driven collapse, are idealized collapse modes and are likely to bracket the true mode of collapse. 

\section{SUMMARY AND CONCLUSIONS} 
\label{conclusion}

We have calculated homologous collapse timescales for the interiors of uniform-density cylinders, based upon a first-order approximation to the accelerations along the major axis of the cylinders. We have also calculated the collapse timescale for a uniform-density cylinder, under the approximation that the central density remains constant, by associating the gravitational force per unit mass on the edge with the rate of change of the momentum per unit mass of the edge. With these results, in conjunction with the homologous collapse timescales of uniform-density circular sheets calculated by J. A. Toal\'{a} et al.\ (2012, in preparation [erratum]) and reproduced in Appendix \ref{2d}, we find that

\begin{itemize}

\item Two separate collapse modes are present within circular sheets and filaments. The interiors of these clouds collapse roughly homologously while the edges are preferentially given more momentum, such that the edges sweep up material and form density enhancements. The effect of preferential edge acceleration has been previously noted in simulations and analytic studies (e.g., \citealt{Bastien83, Burkert04, Hartmann07, VazquezSemadeni07, Heitsch08Slyz, Hsu10, Pon11, Toala12}).

\item The homologous collapse mode is dominant in circular sheets while the edge-driven collapse mode dominates the momentum imparted to filaments with aspect ratios larger than 5.

\item The homologous collapse timescales for the interiors of filamentary clouds scale linearly with $A=Z(0)/{\cal R}$, where $2Z(0)$ is the total initial length and ${\cal R}$ is the radius of the filamentary cloud. The edge-driven collapse mode (constant density collapse) of filamentary clouds produces collapse timescales that are proportional to $\sqrt{A}$. Thus, preferential edge acceleration is most important for clouds with large aspect ratios.

\item Regardless of dimensionality, the acceleration, under the assumption of homologous collapse, can be expressed as
\begin{equation}
\frac{dv_{0}}{dt}\sim  \frac{GM}{\chi^{2}} \nonumber, 
\end{equation}
\noindent with $\chi$ as the collapsing dimension (e.g., radius $R$ for spherical and sheet-like clouds, and the semimajor axis $Z$ for a cylinder). Thus, each reduction of dimension produces an additional $\sqrt{A}$ dependence in the homologous collapse time.

\item In general, lower dimensional objects (``1D'' and ``2D'') and objects with higher aspect ratios have larger collapse timescales than for a sphere with the same volume density.
  
\item Estimates of star formation rates from gas densities can be overestimated by an order of magnitude, for realistic filamentary aspect ratios, if the geometry of a cloud is not taken into account.

\end{itemize}

\acknowledgements A.P. was partially supported by the Natural Sciences and Engineering Research Council of Canada graduate scholarship program. J.A.T. thanks CONACyT, CONACyT-SNI (Mexico) and CSIC JAE-PREDOC (Spain) for a student grant. D.J. acknowledges support from an NSERC Discovery Grant. F.H. gratefully acknowledges support by the NSF through grant AST 0807305 and by the NHSC through grant 1008. E.V.S. acknowledges support from CONACYT Grant 102488. G.C.G. acknowledges support from UNAM-DGAPA grant PAPIIT IN106511. This research has made use of NASA's Astrophysics Data System and the astro-ph archive. We also thank our anonymous referee for many useful changes to this paper.

\bibliographystyle{apj}
\bibliography{ponbib}{}

\newpage 

\appendix

\section{APPENDIX A: HOMOLOGOUS COLLAPSE TIMESCALE OF A SPHERE} 
\label{3d}

The free-fall collapse timescale of a uniform-density sphere is a well-studied problem \citep[e.g.,][]{Binney87} and it is known that the collapse proceeds homologously. For a sphere with a volume density of $\rho$, the acceleration at a distance $r$ from the center of the sphere, $a(r) = 4 \pi G \rho r / 3$, is linearly dependent upon the radial distance, as required for homologous collapse.

The governing differential equation for the collapse of a sphere is
\begin{equation}
\label{eqn:3dode}
\frac{dv_0(t)}{dt} = \frac{G M}{R(t)^2},
\end{equation}
where $v_0(t)$ is the velocity of the edge at time $t$, $R(t)$ is the total radius of the sphere at time $t$, and $M = 4\pi \rho(0) R(0)^3 / 3$ is the total mass of the sphere, with $\rho(0)$ being the volume density at $t=0$. This differential equation is well known and can be solved for the classical free-fall timescale of a sphere, $\tau_{\mathrm{3D}}$,
\begin{eqnarray}
\label{eqn:t3drho}
\tau_\mathrm{3D} &=& \sqrt{\frac{3 \pi}{32G \rho(0)}}.
\end{eqnarray}
	
\section{APPENDIX B: HOMOLOGOUS COLLAPSE TIMESCALE OF A CIRCULAR SHEET}
\label{2d} 

The homologous collapse timescale of a circular sheet is derived in J. A. Toal\'{a} et al.\ (2012, in preparation [erratum]) and the general derivation is reproduced here. We consider a circular sheet with mass $M$, radius $R(t)$, and surface density $\Sigma(t)$, such that the initial radius and surface density are $R(0)$ and $\Sigma(0) = M/ [\pi R(0)^{2}]$, respectively. \citet{Burkert04} show that for an infinitesimally thin circular sheet, the radial acceleration at a distance $r$ from the center of the sheet is
\begin{equation}
a(r,t) = 4G\Sigma(t) \frac{R(t)}{r} \left[K\left(\frac{r}{R(t)}\right) - E\left(\frac{r}{R(t)}\right)\right],
\label{eqn:a2dexact}
\end{equation}
where $K$ is the first complete elliptic integral and $E$ is the second complete elliptic integral. They note that the acceleration of such an infinitely thin sheet is infinite at the edge, where $r = R(t)$, but that in a sheet with finite thickness, the acceleration at the edge is finite.

Due to this infinite acceleration at the edge, it is common (e.g., \citealt{Burkert04, Pon11,Toala12}) to use a first-order approximation to the acceleration,
\begin{equation}
a(r,t) \approx \pi G \Sigma(t) \frac{r}{R(t)}.
\label{eqn:a2dapprox}
\end{equation}
The terms excluded from the above equation are higher order terms of $r / R(t)$, such that the above equation is not valid at, or near, the edge of a circular sheet, but is reasonably accurate for most points within the interior of a circular sheet. 

While we use Equation (\ref{eqn:a2dapprox}) in the following derivation, we assume that the sheet has a constant, finite height of $H$ that is much smaller than the initial radius $R(0)$, such that we can define the aspect ratio as $A = R(0) / H$. We thus rewrite Equation (\ref{eqn:a2dapprox}) as
\begin{equation}
  a(r, t) \approx \pi G \rho(t) H \frac{r}{R(t)},
\label{eqn:a2drho}
\end{equation}
where $\rho(t) = \Sigma(t)/H$ is the volume density of the sheet at time $t$.

It is critical to note that the acceleration in Equation (\ref{eqn:a2drho}) is a linear function of the radius and, thus, this equation exactly describes a homologous collapse. That is, to first order, the accelerations of an infinitely thin, circular sheet will cause the sheet to collapse homologously. It is only the higher order terms of $r / R$ that cause a deviation from homologous collapse.

Denoting the velocity at the edge as $v_0(t)$ and using the relation between the mass of a circular sheet and its volume density, $M = \pi R^3 \rho / A$, Equation (\ref{eqn:a2drho}) can be re-written as
\begin{equation}
\label{eqn:2dode}
\frac{dv_0}{dt} \approx \frac{G M}{R(t)^2}.
\end{equation}
Equation (\ref{eqn:2dode}) can be solved to show that the collapse timescale of a circular sheet, $\tau_{\mathrm{2D}}$, is
\begin{eqnarray}
\label{eqn:t2drho}
\tau_\mathrm{2D} &\approx& \sqrt{\frac{4 A}{3}} \, \tau_\mathrm{3D}.
\end{eqnarray}
Thus, under the assumption of homologous collapse, the timescale for collapse of a circular sheet scales with $\sqrt{A}$.

\section{APPENDIX C: UNIFORM-DENSITY COLLAPSING FILAMENT}
\label{app:filament dens} 

Assuming that the only force on the edge of a cylinder is from the uniform-density interior material, this gravitational force from the interior must be equal to the rate of change of the momentum of the edge. From Newton's second law,
\begin{equation}
m(t) g = v(t) \frac{dm(t)}{dt} + m(t) \frac{dv(t)}{dt},
\end{equation}
where $v(t)$ is the velocity of the edge and $m(t)$ is the mass of the edge. We define the sign of the velocity such that $v(t)$ is positive for inward motions. See also Section \ref{uniform density} for other variable definitions. Since the mass at the edge increases as material is swept up
\begin{equation}
\frac{dm(t)}{dt} = \pi {\cal R}^2 \rho \, v(t).
\label{eqn:dmdt1d}
\end{equation}
Thus, the differential equation governing the motion of the edge is
\begin{equation}
m(t) g = \pi {\cal R}^2 \rho \, v(t)^2 + m(t) \frac{dv(t)}{dt}.
\label{eqn:1dpdedens}
\end{equation}

For the remainder of this derivation, we drop the functional dependence on $t$ from our notation. Rewriting the derivative of the velocity yields
\begin{eqnarray}
\frac{dv}{dt} &=& \frac{dv}{dm} \frac{dm}{dt},\\
\frac{dv}{dt} &=& \frac{dv}{dm} \pi {\cal R}^2 \rho \, v,\\
m g &=& \pi {\cal R}^2 \, \rho \, v^2 +m \frac{dv}{dm} \pi {\cal R}^2 \, \rho \, v,\\
\frac{dv}{dm} v &=& \frac{-v^2}{m} + \frac{g}{\pi {\cal R}^2 \, \rho}.
\end{eqnarray}

This is an Abel differential equation of the second kind. It can be solved with the aid of the substitution
\begin{eqnarray}
v &=& \frac{w}{m},\\
\frac{dv}{dm} &=& \frac{dw}{dm} \frac{1}{m} - \frac{w}{m^2},\\
\frac{dw}{dm} \frac{w}{m^2} &=& \frac{g}{\pi {\cal R}^2 \, \rho},\\
\frac{w^2 - w(t = 0)^2}{2} &=& \frac{\left(m^3 - m(t = 0)^3\right) \, g}{3\pi {\cal R}^2 \, \rho}.
\end{eqnarray}
At the beginning of the collapse, $m = 0$ and $v = 0$, such that $w(t = 0 ) = 0$. We have also used the fact that the gravitational force per unit mass, to lowest order, is constant over the collapse.

The edge velocity, as a function of the edge mass, is thus
\begin{equation}
v^2 = \frac{2m \, g}{3\pi {\cal R}^2 \, \rho}.
\label{eqn:vofm1d}
\end{equation}

Since the mass at the edge is equal to the mass swept up,
\begin{eqnarray}
m &=& (Z(0) - Z) \pi {\cal R}^2 \rho,\\
-\frac{dZ}{dt} &=& \sqrt{\frac{2(Z(0) - Z) \, g}{3}},\\
t &=& \sqrt{\frac{6(Z(0) - Z)}{g}},\\
Z &=& Z(0) - \frac{g \, t^2}{6}.
\end{eqnarray}

\end{document}